\documentclass[journal ]{new-aiaa}
\usepackage[utf8]{inputenc}
\usepackage{textcomp}

\usepackage{graphicx}
\usepackage{amsmath}
\usepackage[version=4]{mhchem}
\graphicspath{{./Figures/}}
\usepackage[separate-uncertainty=true,multi-part-units=single]{siunitx} 
\usepackage{booktabs}
\usepackage{longtable,tabularx}
\title{Aerodynamic drag reduction of a tilt rotor aircraft using zero-net-mass-flux devices}

\author{Hung Truong \footnote{Postdoc, truongdinhhung90@gmail.com}, Abderahmane Marouf \footnote{Associate Professor, amarouf@unistra.fr} and Yannick Hoarau \footnote{Professor, hoarau@unistra.fr}}
\affil{Mechanical Department, ICUBE laboratory, University of Strasbourg, 67000, Strasbourg, France}
\author{Jan B. Vos \footnote{Director, jan.vos@cfse.ch, Senior AIAA member} and Alain Gehri \footnote{Senior Scientist, alain.gehri@cfse.ch}}
\affil{CFS Engineering, EPFL Innovation Park, Batiment A, CH-1015 Lausanne, Switzerland}

\begin{document}

\maketitle

\begin{abstract}
The study described in this paper was conducted as part of the European Funded CleanSky2 project AFC4TR (Active Flow Control for Tilt-Rotor aircraft).
High-Fidelity numerical simulations were made to study various approaches of using Active Flow Control (AFC) actuators to delay flow separation at near stall conditions of the Next Generation Civil Tilt Rotor (NGCTR) VTOL aircraft with a tilted nacelle used during take-off and landing. The study revealed that for this configuration the flow separations travel in the spanwise direction starting from the middle of the wing. Various flow control strategies using Zero Net Mass Flux (ZNMF) actuators (synthetic jets) were then tested. Different number of actuators were integrated on the wing at different positions in order to optimize the effectiveness of these devices. It was found that when placed correctly, using the optimal blowing velocity and actuation frequency, ZNMF devices help to delay flow separation, resulting in a reduction of pressure drag and an increase in the aircraft's aerodynamic efficiency.
\end{abstract}

\section*{Nomenclature}

{\renewcommand\arraystretch{1.0}
\noindent\begin{longtable*}{@{}l @{\quad=\quad} l@{}}
$C_{ref}$ & referenced chord (\si{\meter}) \\
$D$ &  jet diameter (\si{\meter})\\
$CD$ & drag coefficient \\
$CL$ & lift coefficient \\
$CF$ & skin friction coefficient \\
$CP$& pressure coefficient \\
d$t$ & time step (\si{\second}) \\
$U_{\infty}$ & streamflow velocity (\si{\meter\per\second}) \\
$F^+$ & dimensionless frequency of the jet \\
$f_{jet}$ & blowing frequency of the jet (\si{\hertz}) \\
$T$ & blowing period of the jet (\si{\second}) \\
$V_{jet}$ & maximum blowing velocity of the jet (\si{\meter\per\second}) 
\end{longtable*}}

\section{Introduction}\label{sec1}

Between 2020 and 2022, the aviation industry worldwide has been impacted significantly by the Covid pandemic, but as travel restrictions have been lifted, air travel is approaching pre-pandemic levels and is expected to continue to grow over the next 20 years. The aviation sector faces a major challenge of reducing its environmental impact. To address this challenge, the Advisory Council for Aviation Research and Innovation in Europe has set goals to reduce emissions, noise and fuel burn by 90\%, 65\% and 75\% respectively by 2050.

The Next Generation Civil Tilt Rotor (NGCTR) is a new concept that combines the advantages of helicopters and fixed-wing aircrafts. It has a similar range and speed as a fixed-wing aircraft, but can take off and land vertically and hover when necessary. A technology demonstrator will be built to test its design, systems and operational concepts, including the use of Active Flow Control to increase aerodynamic efficiency and reduce fuel consumption, emissions and $CO_2$. Despite being studied for over 50 years, Active Flow Control has not been widely adopted due to integration and power supply issues, reliability, and certification requirements. A large number of AFC systems are available on the market, and they might be classified according their input energy, orientation with respect to the external flow, frequency response and bandwidth, \cite{Cattafesta2011}. This paper is concerned with the use of so called Zero-Net-Mass-Flux (ZNMF) actuators, which belong to the category of fluidic devices. The design of a tilt rotor with ZNMF devices from the start may lead to a wider use of these devices for future and current aircraft.

There have been notable advancements in the field of active flow control using ZNMF devices on airfoils in recent years. Seifert and Pack~\cite{Seifert99_highRe} were among the first to experiment active flow control on a NACA0015 airfoil at high Reynolds number. They investigated the usefulness of synthetic jets and the results were supported by subsequent experiments~\cite{McCormick00_directedSJ,Tuck04_ZNMF,TANG14_SJA} that showed the benefits of ZNMF devices for increasing lift and reducing drag. In parallel, many researchers used numerical methods to examine the impact of synthetic jets on different airfoils as in~\cite{Donovan97_Numerical, YOU08_NumLES, KIM09_Numerical}, just to name a few. These studies generally concluded that placing the jets upstream of the separation point helped to delay flow detachment, and that the effectiveness of the jets was proportional to the velocity ratio. Optimizing the parameters of synthetic jets is computationally challenging and has been the subject of only a few studies. Duvigneau and Visonneau~\cite{DUVIGNEAU06_NumOpt} combined an optimization process with a fluid solver to find the best performance for a synthetic jet located 12\% of the way from the leading edge of a NACA0015 airfoil. Montazer et al.~\cite{Montazer16_NumOpt} optimized the position and frequency of a synthetic jet using the Response Surface Methodology. More recently, the authors investigated two-dimensional active control using different blowing strategies similar to piezo-actuators embedded in the flap of a high-lift system~\cite{Marouf2021FSSIC,BmegaptcheTekap2021,BmegaptcheTekap2019}. The active flow control was combined with a morphing flap through cambering, which led to large flow separations on the suction side of the flap and a decrease in aerodynamic efficiency. However, the study showed that coupling active flow control with morphing could reattach the boundary layer, improve the pressure distribution over the flap and wing, and decrease the size of the wake.

While previous research has shown the advantages of using ZNMF devices on airfoils, the use of ZNMF devices on a whole aircraft has rarely been studied, especially for a tilt-rotor aircraft. In the present work, this gap in knowledge is
filled through the use of high fidelity CFD simulations of unsteady flow. The role of AFC will be examined for the three-dimensional NGCTR aircraft in the context of the Active Flow Control for Tilt-Rotor aircraft (AFC4TR), a CleanSky2 European funded project (2020-2022). This paper is organized as follows. In section 2, the numerical methods including the CFD solver, the meshing and the ZNMF modeling will be presented. Then the effectiveness of the ZNMF devices in delaying flow separation on the wing surface will be discussed for different configurations in section 3. Finally, conclusions are given along with  possible future work in section 4.

\section{Numerical Approach}

\subsection{Navier Stokes Multi Block (NSMB) solver}

The work presented in this paper is made using the NSMB CFD solver.
NSMB is a parallelized CFD solver that uses the finite volume method to solve the Reynolds-averaged Navier Stokes equations for compressible flows on multi-block structured grids. The solver, developed by a consortium of universities and industries, has been used for aerospace applications for over 30 years \cite{Vos1998}. It includes features such as ALE~\cite{Donea1982}, turbulence models~\cite{Hoarau2002}, Chimera overlapping~\cite{Hoarau2016}, chemistry modeling, grid flexibility, and FSI coupling for morphing wing modeling~\cite{Marouf2020,Simiriotis2019,T2019}. Space discretization schemes include central schemes and upwind schemes up to 5th order. Time integration methods include explicit and implicit schemes, and convergence acceleration techniques can be used. NSMB also has various validated turbulence models like Spalart-Allmaras~\cite{spalart1992one}, Organized Eddy Simulation~\cite{Hoarau2002,Bourguet2008}, and two-equation by Menter~\cite{Menter93}. Unsteady simulations can be conducted using dual time stepping or Newton's approach.

\subsection{Computational setup and mesh generation}

\begin{figure}[hbt!]
\centering
\includegraphics[width=.45\textwidth]{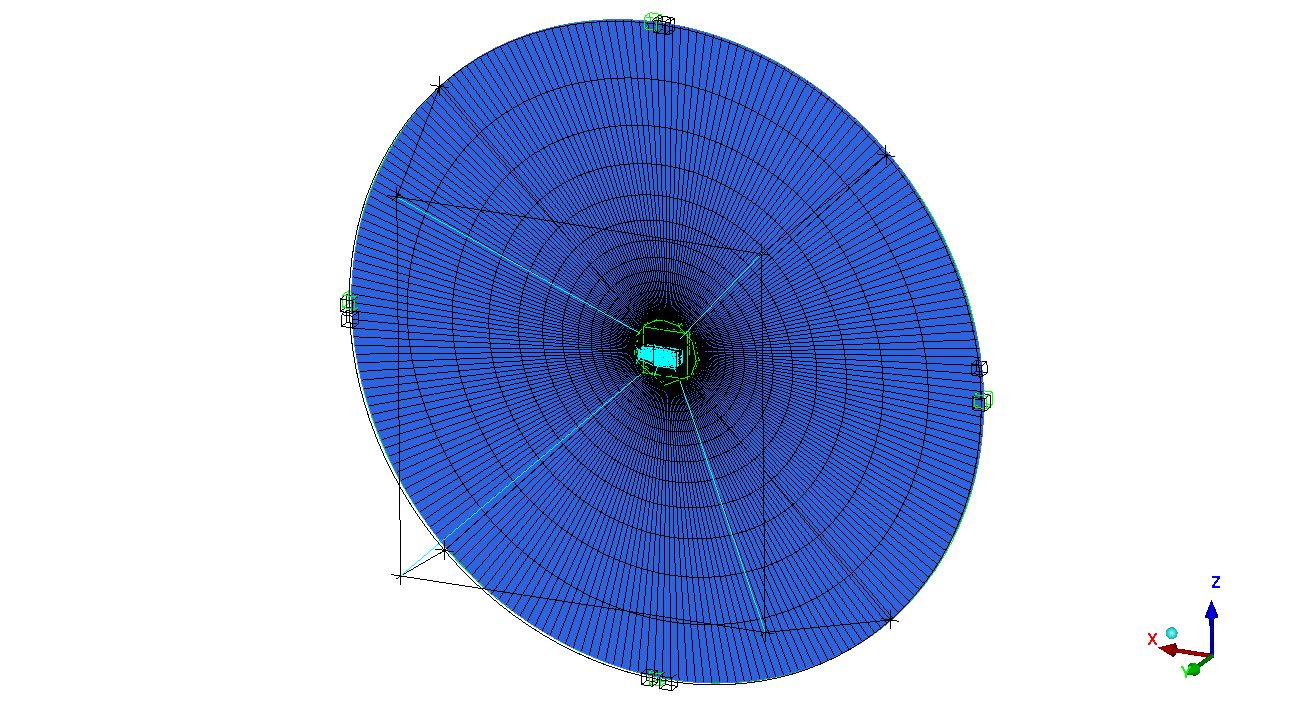}
\includegraphics[width=.45\textwidth]{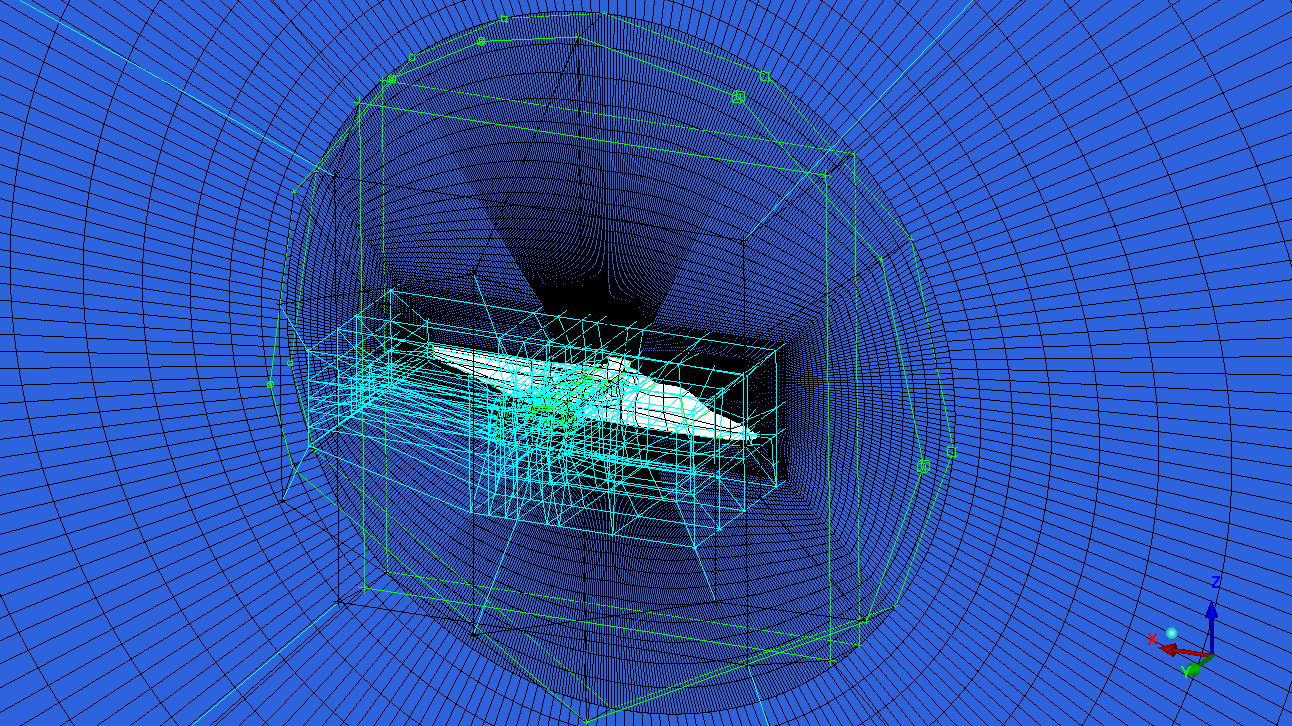}
\caption{Patch grid connectivity between far field (left) and near field around the aircraft (right). Topology of the grid with far-field boundary at $100 \times C_{ref}$.}
\label{fig:grid_topology}
\end{figure}

\begin{figure}[hbt!]
\centering
\includegraphics[width=0.85\textwidth]{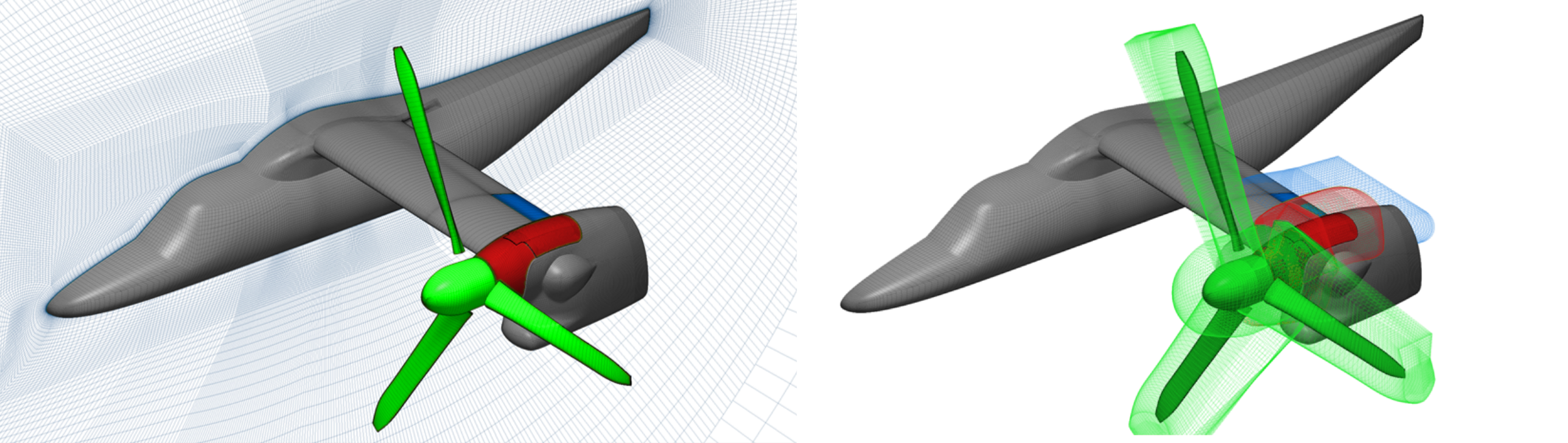}
\caption{Left: surface mesh around the aircraft and on the symmetry boundary. Right: volume mesh around chimera components: rotating propeller (green), tilted nacelle (red) and deflected aileron (blue).}
\label{fig:mesh}
\end{figure}

The NGCTR aircraft was studied at near stall conditions with the inflow condition of $Re=4.4 \times 10^6$. The multi-block structured grids required by the NSMB flow solver were generated by the ANSYS ICEM CFD pre-processor tool. The gridding guidelines of the AIAA CFD High Lift/Drag Prediction Workshops were followed for the height of the initial cell size near walls, and for the cell growth ratio normal to the walls. The patch grid connectivity was used to decrease the number of cells in the far-field region, and the far-field boundary was placed at 100 times the reference wing chord $C_{ref}$ as shown in Fig.~\ref{fig:grid_topology}. 
The Chimera overlapping grid technique was used to simplify the mesh generation process by creating independent grids for parts that can move (as for example the rotor, nacelle and aileron). The final grid is then created from the independent parts, making the appropriate rotations and/or translations when needed.
Fig.~\ref{fig:mesh} shows the mesh on the airplane surface as well as the volume mesh around chimera components. In order to accurately model the viscous boundary layer, O-grid topologies were used near solid surfaces such as the fuselage, wing, aileron, nacelle, engine, and rotor blades. The first cell height in the wall normal direction was chosen to obtain a $y+$ value of around 1, ensuring the use of low-Reynolds turbulence modeling. The cell size $\Delta y$ in the wall normal direction was kept below $\SI{0.006}{\milli\meter}$ and the growth ratio of the cells normal to the wall was about 1.2. The chordwise spacing was set to be below 0.1\% of the local chord at the leading and trailing edges of the main wing and the rotor blades. On the other hand, the spanwise spacing was set to be below 0.1\% of the semi-span at the root (fuselage junction) and at the tip (nacelle junction).

\subsection{Modeling of ZNMF actuators}

The ZNMF actuators produce an oscillating jet either by an sinusoidal oscillating membrane or a piston to alternating force fluid through an orifice into the external flow field and entrain fluid back. During the blowing stroke a vortex will be produced at the orifice and is entrained with the external flow. The efficiency of ZNMF actuators depends on various parameters of which the two key ones are the dimensionless frequency of the jet $F^+=f_{jet} C_{ref} / U_{\infty}$ and the ratio of the actuator velocity to the free-stream velocity $V_{jet} / U_{\infty}$ (or the blowing momentum coefficient $C_{\mu} = V_{jet}^2 D / \frac{1}{2} U_{\infty}^2 C_{ref}$). Prior studies~\cite{Truong2023} have been performed to investigate the influence of the jet velocity and frequency. Consequently, the maximum blowing velocity and the jet frequency are fixed at $\SI{300}{\meter\per\second}$ and $\SI{65}{\hertz}$, respectively.  

\begin{figure}[hbt!]
\centering
\includegraphics[width=0.75\textwidth]{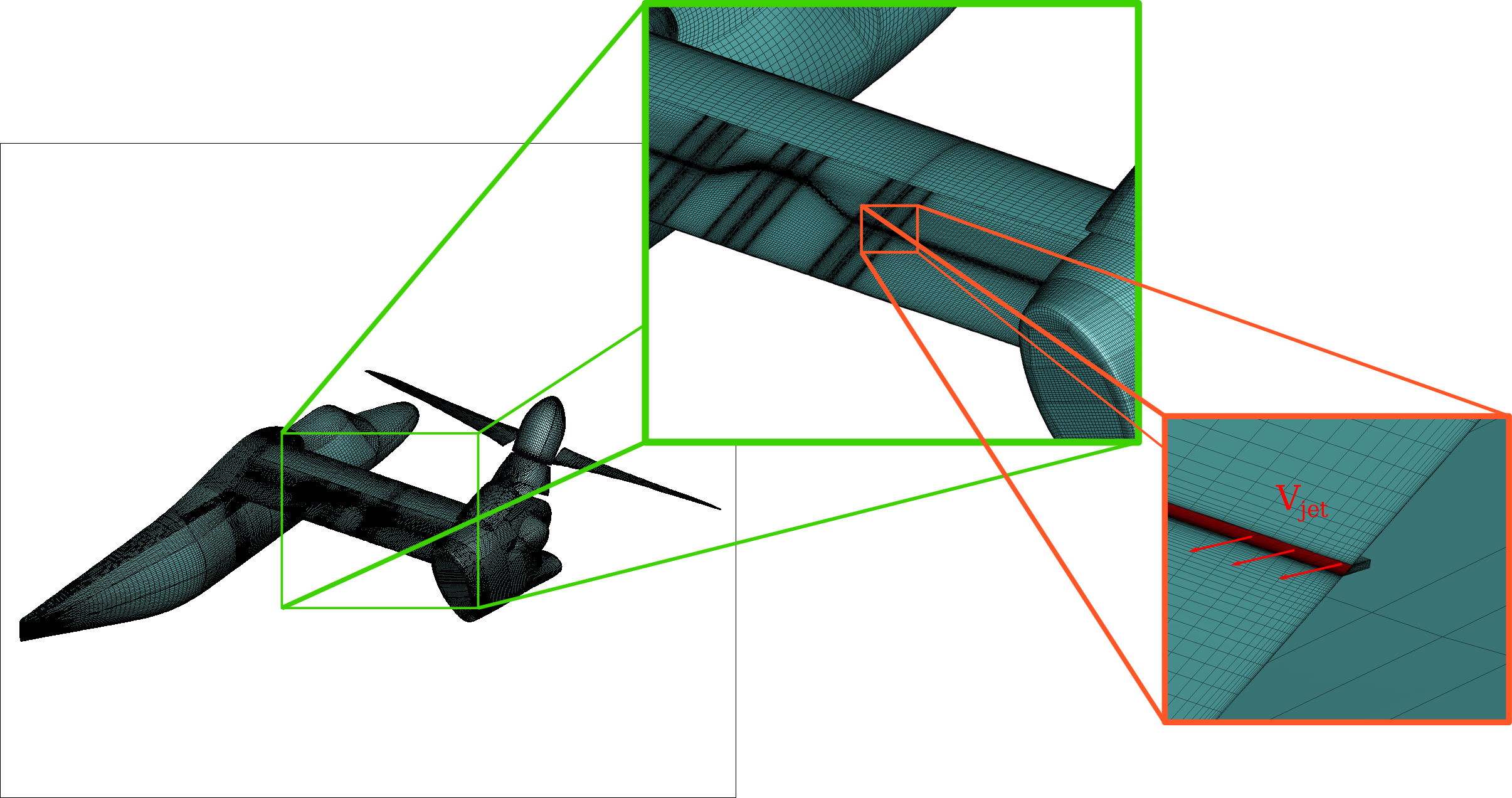}
\caption{ZNMF devices embedded inside the wing as rectangular holes. The mesh in the vicinity of the ZNMF was refined in order to well resolve the sucking and blowing jet.}
\label{fig:grid_ZNMF}
\end{figure}

From a modeling point of view, two distinct approaches have been suggested: one involves modeling the velocity of the jet, while the other involves modeling the membrane that triggers the motion. The former was chosen for the study due to its simplicity and efficiency. In addition, the chimera approach allows to implement easily AFC actuators in the aircraft's wing without any remeshing and to facility the optimization study of the ZNMF position. Particular attention was also paid to the refinement of the mesh close to the areas where the ZNMF devices will be placed later in the study (cf. Fig.~\ref{fig:grid_ZNMF}). The dimension of the jets is similar to the one of a high efficiency innovative synthetic jet actuator using Amplified Piezo-Actuator developed in the context of the French National project ASPIC~\cite{Eglinger18_ASPIC}. More 
specifically, it has a width of \SI{150}{\milli\meter}, an exit hole of
\SI{1}{\milli\meter} and an angle of \SI{45}{\degree} oriented in the 
same direction as the external flow.

\subsection{Time step and tolerance}

\begin{figure}[hbt!]
\centering
\includegraphics[width=0.75\textwidth]{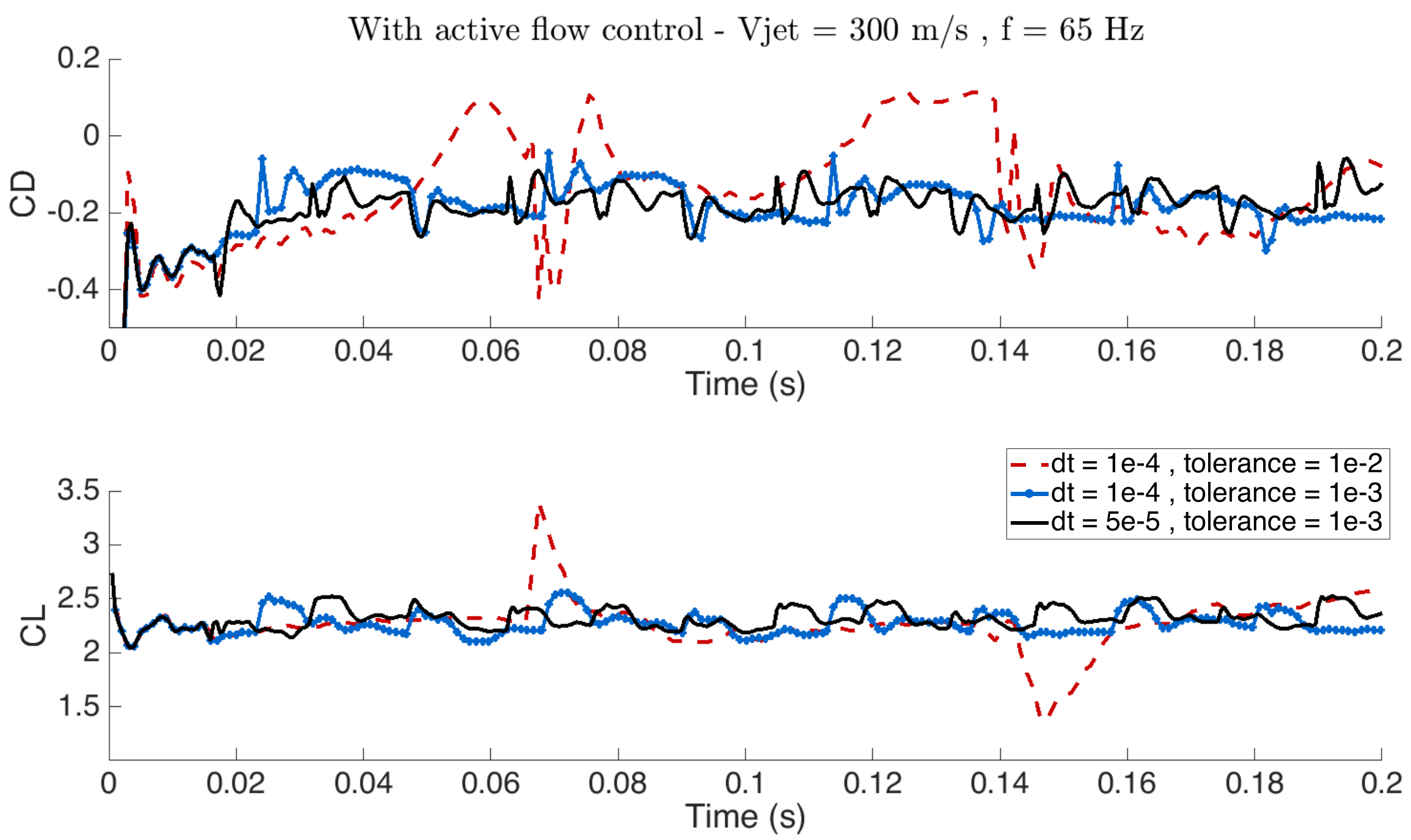}
\caption{Drag and lift coefficient evolution over time for different time steps and tolerances.}
\label{fig:timestep_CLCD}
\end{figure}

\begin{figure}[hbt!]
\centering
\includegraphics[width=0.55\textwidth]{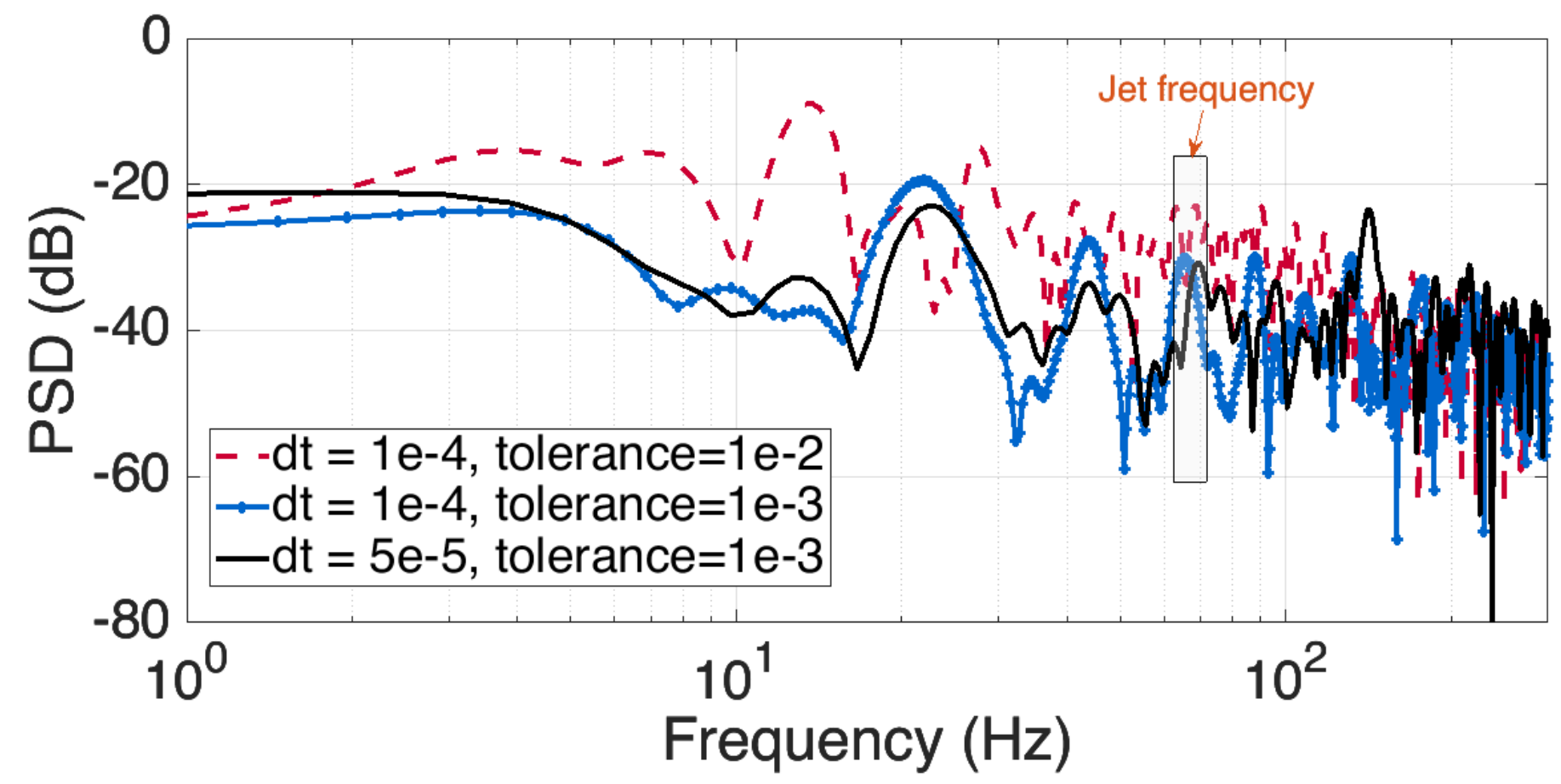}
\caption{Power Spectral Density of the drag coefficient for different time steps and tolerances.}
\label{fig:timestep_PSD}
\end{figure}

Prior to the detailed studies of AFC using ZNMF devices, the two most important numerical parameters for the dual time stepping scheme, i.e. the time step and the tolerance, were investigated. The drag and lift coefficients were plotted for different time steps and minimum fixed tolerances (as shown in Fig.~\ref{fig:timestep_CLCD}). The maximum blowing velocity of the ZNMF jets was set to \SI{300}{\meter\per\second} and the actuation frequency at \SI{65}{Hz}, and all simulations were conducted using a 3-bladed rotating propeller. It was found that with a time step of $dt = \SI{1e-4}{\second}$ and a tolerance of $1e-2$, non-physical peaks were present in both the drag and lift signals. However, when the tolerance was reduced to $1e-3$ and the time step was fixed at $1e-4$ and $5e-5$ seconds, these peaks were no longer overestimated. Both cases with the same tolerance but different time steps showed similar results. 
The Power Spectral Density (PSD) of the drag coefficient is shown in Fig.~\ref{fig:timestep_PSD}, and both cases with $dt = \SI{1e-4}{\second}$ \& tolerance $1e-3$ and $dt = \SI{5e-5}{\second}$ \& tolerance $1e-3$ captured the predominant frequency of the rotating blades and the frequency of the jet. 
Consequently, the case with a time step of $dt = \SI{1e-4}{\second}$ and a tolerance of $1e-3$ was used for the simulations of the tilt-rotor with the ZNMF devices because it presented a good compromise between maintaining sufficient accuracy and reasonable computational time.

\begin{figure}[hbt!]
\centering
\includegraphics[width=0.85\textwidth]{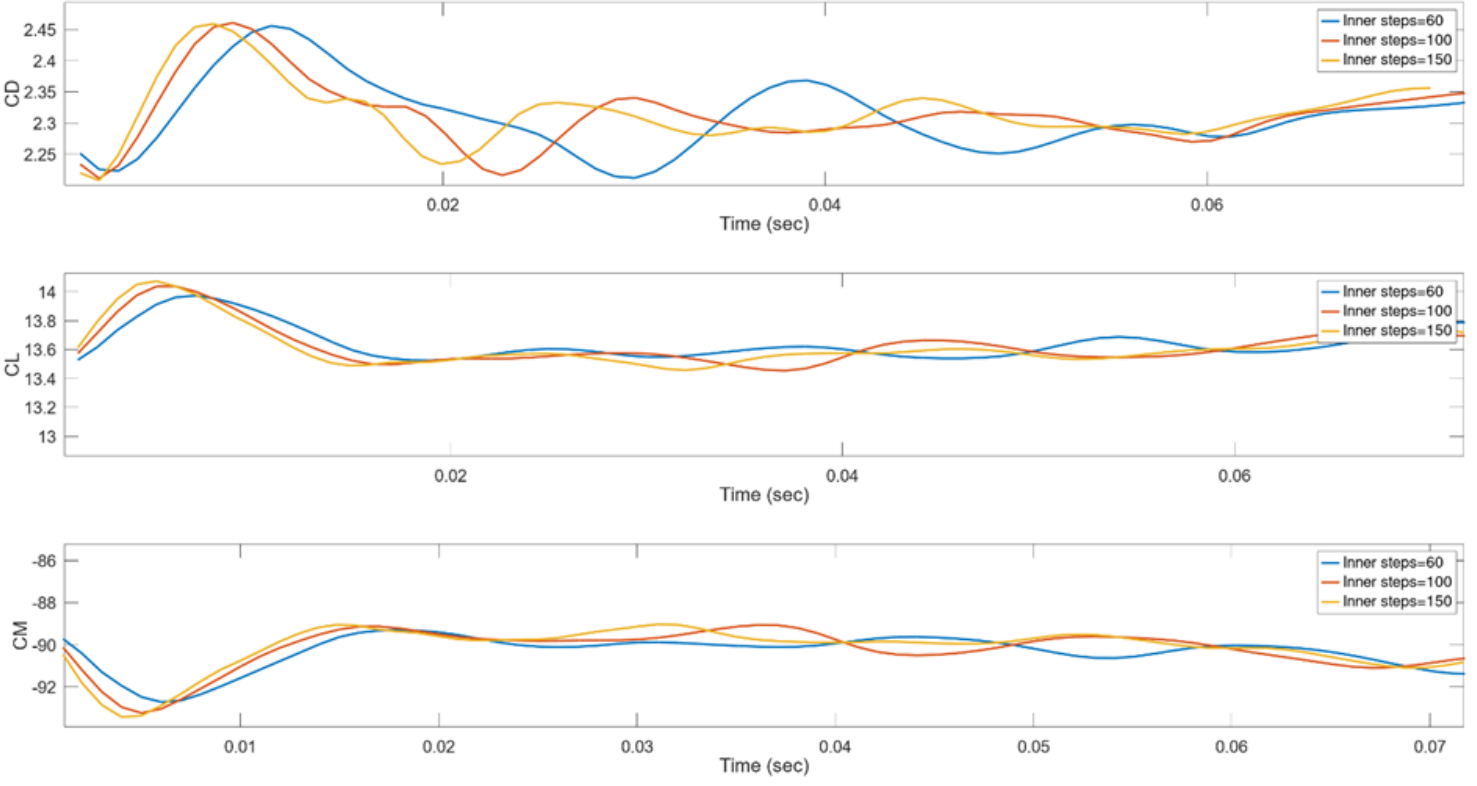}
\caption{Comparison of unsteady simulations of the aircraft using the implicit dual time stepping scheme with different inner steps (loop of convergence)}
\label{fig:timestep_innerstep}
\end{figure}

Another important parameter of the dual time stepping scheme is the maximum number of inner steps used to converge the solution in the inner loop in case the tolerance criterium is not met. Calculations were made using the outer time step of $dt = \SI{1e-4}{\second}$ and a tolerance of $1e-3$ for inner steps $nsteps = 60, 100$ and $150$. It was found that setting the number of inner steps to $100$ is a good compromise, see also Fig.~\ref{fig:timestep_innerstep}.

\section{Results and Discussions}\label{sec2}

\subsection{Baseline flow}

\begin{figure}[hbt!]
\centering
\includegraphics[width=0.67\textwidth]{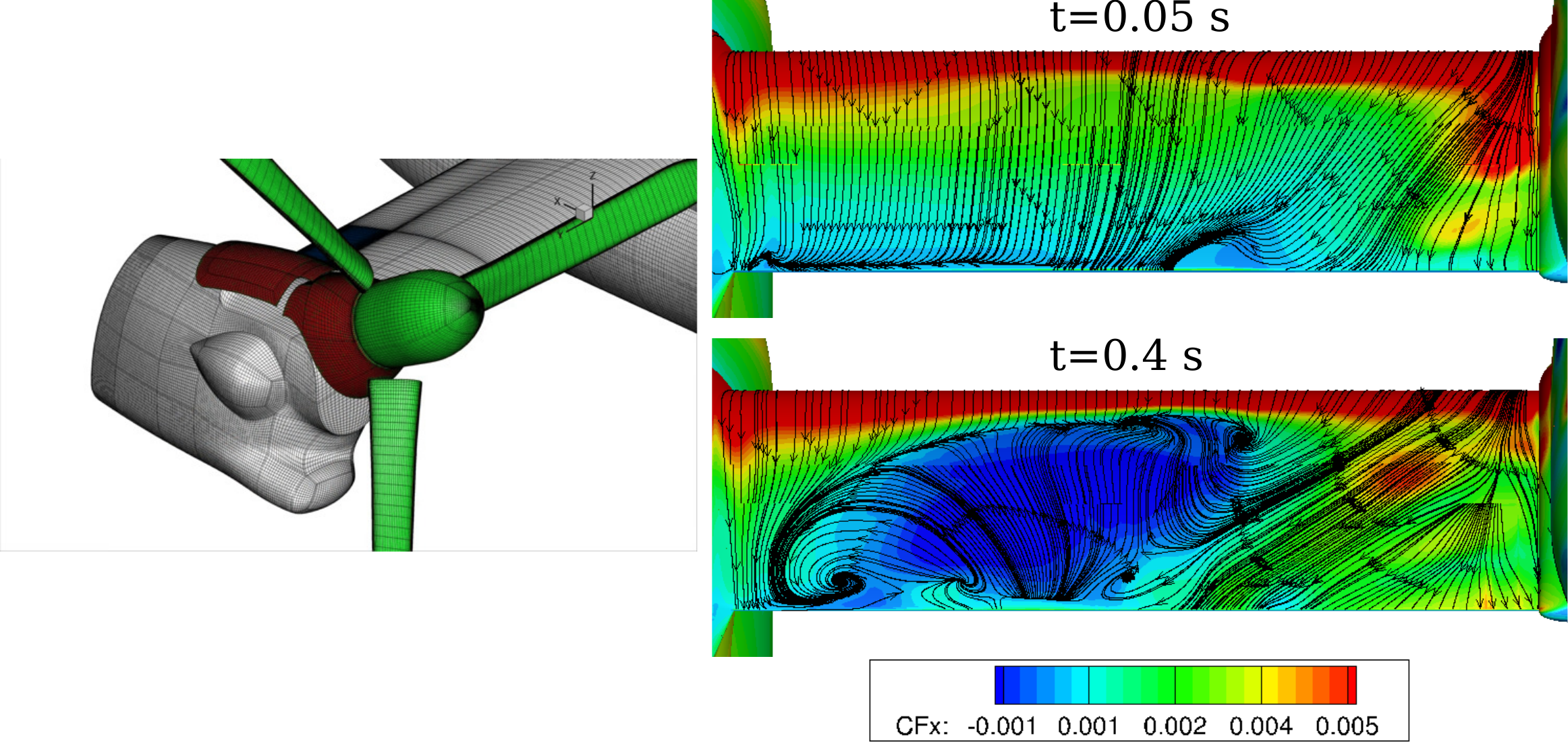}
\caption{Left: The studied configuration of the NGCTR aircraft at high angle of attack with a tilted nacelle can benefit from active flow control using ZNMF jets. Right: Skin friction $CF_x$ component contours with streamlines of CF ($CF_x, CF_y, CF_z$), for the baseline flow at two different time instant.}
\label{fig:baseline_config2}
\end{figure}

The aim of the study was to investigate the potential benefits of using ZNMF devices for reducing flow separations at high angle of attack. To this end, unsteady simulations were firstly conducted around the Leonardo Helicopters NGCTR aircraft to identify critical flow configurations that caused boundary layer detachment leading to a decline in aerodynamic performance. Particular attention was given to the take-off and landing phases in the subsonic speed range. After conducting numerous simulations, the high angle of attack configuration with a tilted nacelle, as shown on the left of Fig.~\ref{fig:baseline_config2}, was identified as being capable of benefiting from the use of ZNMF devices.

The impact of the configuration with tilted nacelle was investigated with the numerical parameters mentioned previously. It was found that corner vortices are growing near the fuselage-wing and wing-nacelle junctions when the aircraft is operating with a horizontal nacelle~\cite{Truong2023}. Nevertheless, when the nacelle is tilted upwards, the flow is automatically reattached at the wing-nacelle junction since the blockage effect caused by the rotors is reduced and a deviation in the flow begins to form near the nacelle-wing junction. Moreover, at around $60\%$ of the span (distance from the fuselage), a small vortex started forming at the very beginning as shown on the top right of Fig.~\ref{fig:baseline_config2}. The growth of the vortex was caused by the interaction of the rotating blades and the diverted flow. At $0.4 seconds$, the boundary layer is mostly detached and significant flow separation can be observed, which is indicated by the blue area on the bottom right of Fig.~\ref{fig:baseline_config2}. This separation occurs on half of the wing surface and is most noticeable near the trailing-edge, where the separation line is moving in the upstream direction relative to the wing. The simulation revealed that when the nacelle was inclined and the speed of the blades was increased, the flow became unstable and a significant separation of the boundary layer took place. As a result, it is believed that this critical flow would be ideal for using AFC devices.

\subsection{Influence of the ZNMF position}

\begin{figure}[hbt!]
\centering
\includegraphics[width=.35\textwidth]{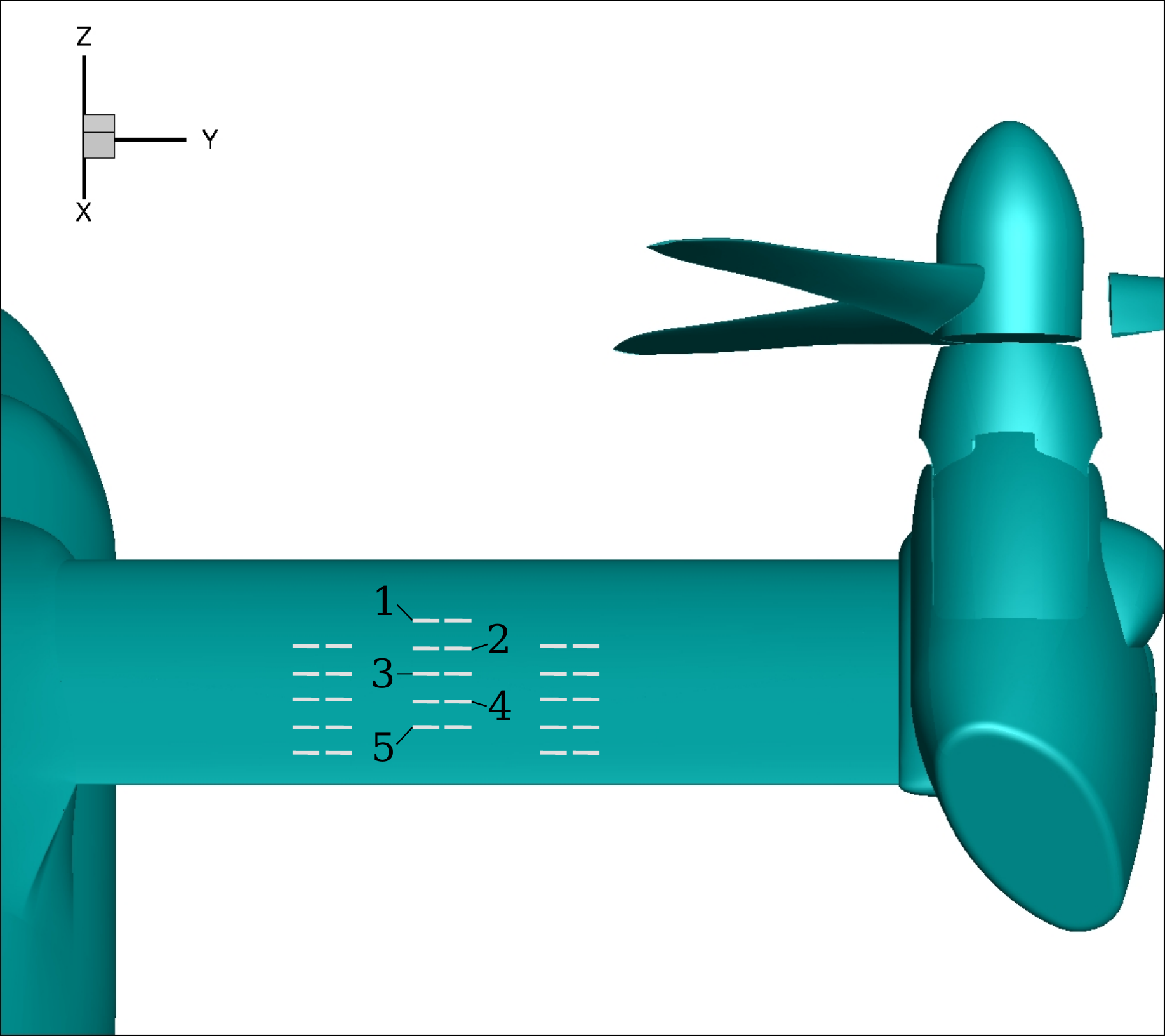}
\caption{Position of the ZNMF devices proposed for active flow control. Six ZNMF devices were used and the position, numbered from 1-5, was varied along the chordwise direction.}
\label{fig:ZNMF_positions}
\end{figure}

Based on the information obtained from the simulations of the baseline flow, different positions of the ZNMF jets were proposed as shown in Fig.~\ref{fig:ZNMF_positions}. For each position, six actuators were integrated at the center of the wing span where the two ones in the middle are placed a little bit upstream comparing to the others. With this setup, the ZNMF devices were expected to be installed upstream of the separation zone in order to control effectively the flow separation. However, how far upstream they should be placed remained an open question. Consequently, 5 different positions in the chordwise direction were considered: the middle position (position 3), upstream $20~\%$ (position 1) and $10~\%$ (position 2) then downstream $10~\%$ (position 4) and $20~\%$ (position 5) of the wing chord.

\begin{figure}[hbt!]
\centering
\includegraphics[width=1.\textwidth]{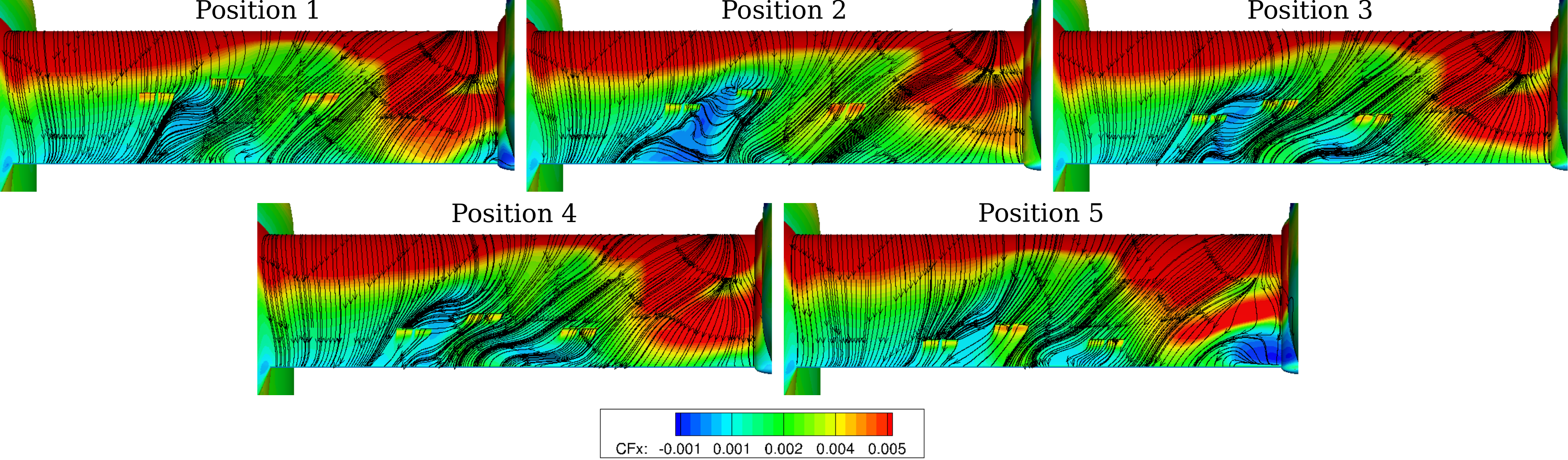}
\caption{Instantaneous skin friction $CF_x$ component contours with streamlines of CF ($CF_x, CF_y, CF_z$) for the baseline (top left) and the controlled cases of configuration 2 with a tilted nacelle and rotating blades.}
\label{fig:case2_CFx_t0428}
\end{figure}

The instantaneous streamwise component of the skin friction $CF_x$ of all the computed cases are shown in Fig.~\ref{fig:case2_CFx_t0428}. When the ZNMF jets are operating, the vortex formed in the middle of the span in the baseline case is effectively suppressed and the flow is reattached. However, this picture is only a snap shot in time.

\begin{figure}[h!]
\centering
\includegraphics[width=0.8\textwidth]{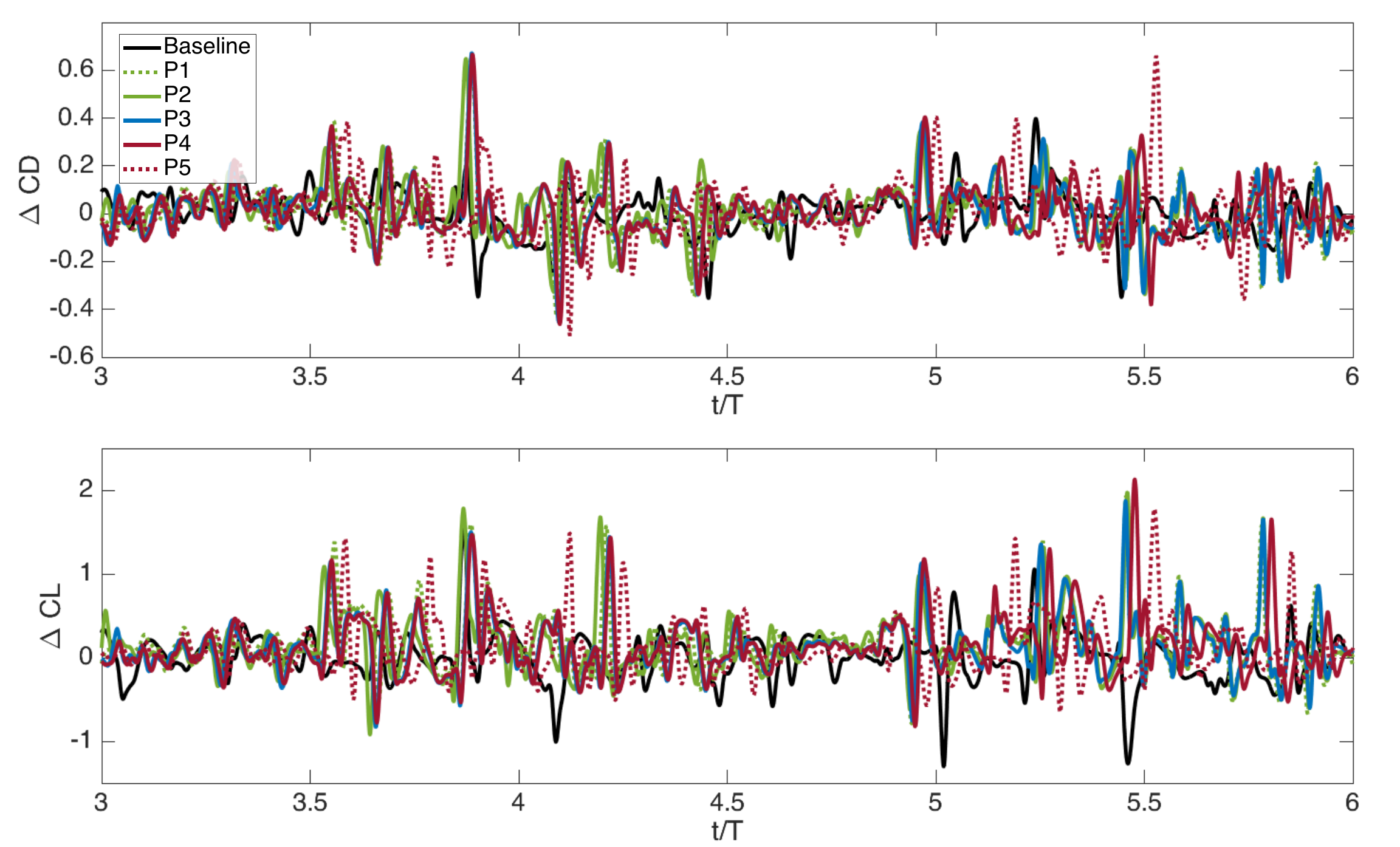}
\caption{Time history of the variation of the drag coefficient $\Delta CD$ (top) and the lift coefficient $\Delta CL$ (bottom) acting on the wing for different positions of the ZNMF with respect to the time-average forces of the baseline flow.}
\label{fig:case2_forces}
\end{figure}

In order to better quantify the impact of the ZNMF actuators, the lift and the drag generated by the wing are then investigated. Due to the influence of the tilted propeller, the forces fluctuates in a wide range. Thus, the average in time of the force coefficients $\widebar{CD}$ and $\widebar{CL}$ were calculated by the integral:

\begin{equation}
    \widebar{CD} = \frac{1}{t_2 - t_1} \int_{t_1}^{t_2} CD(t) dt; \ \ \widebar{CL} = \frac{1}{t_2 - t_1} \int_{t_1}^{t_2} CL(t) dt
\end{equation}

where $t_1$ and $t_2$ were the beginning and the end of the observed period shown on Fig.~\ref{fig:case2_forces}. Then, due to confidentiality reasons, all the forces are shown in the form of the variations with respect to the mean value of the baseline flow:

\begin{equation}
    \Delta CD (t) = CD (t) - \widebar{CD}_{baseline} \ \ \Delta CL (t) = CL (t) - \widebar{CL}_{baseline}
\end{equation}

where $\widebar{CD}_{baseline}$ and $\widebar{CL}_{baseline}$ are the time-average drag and lift of the baseline configuration without ZNMF devices. The time history of the force variation along with the corresponding time-average values are calculated and shown in Fig.~\ref{fig:case2_forces} and Table~\ref{tab:forces_position}. The use of active flow control using ZNMF jets resulted in a significant improvement in lift performance, with an increase of $10.4\%$ to more than $13.5\%$ in lift observed across all positions. This, however, comes at the cost of a significant increase in drag compared to the baseline configuration. In light of this, only the first position, where the jets are placed the furthest upstream, delivers a better aerodynamic efficiency with an increase of $2.91\%$ in the lift-to-drag ratio.

\begin{table}[h]
\begin{center}
\begin{minipage}{174pt}
\caption{The variation of the time-averaged force coefficients with respect to the baseline case for different positions of the ZNMF jets. The values are integrated for the 4th, 5th and 6th rotations of the propeller.}\label{tab:forces_position}%
\begin{tabular}{@{}l || c c c@{}}
\toprule
           & $\Delta C_L$ & $\Delta C_D$ & $\Delta (C_L/C_D)$ \\
\hline \\[-2.0ex]
Position 1 & 10.31\% & 13.51\% & 2.91\%  \\
Position 2 & 18.92\% & 11.93\% & -5.88\% \\
Position 3 & 14.23\% & 11.06\% & -2.77\% \\
Position 4 & 17.68\% & 11.14\% & -5.56\% \\
Position 5 & 18.12\% & 10.41\% & -6.53\% \\
\toprule
\end{tabular}
\end{minipage}
\end{center}
\end{table}

\subsection{Influence of the number of ZNMF devices}

\begin{figure}[hbt!]
\centering
\includegraphics[width=.75\textwidth]{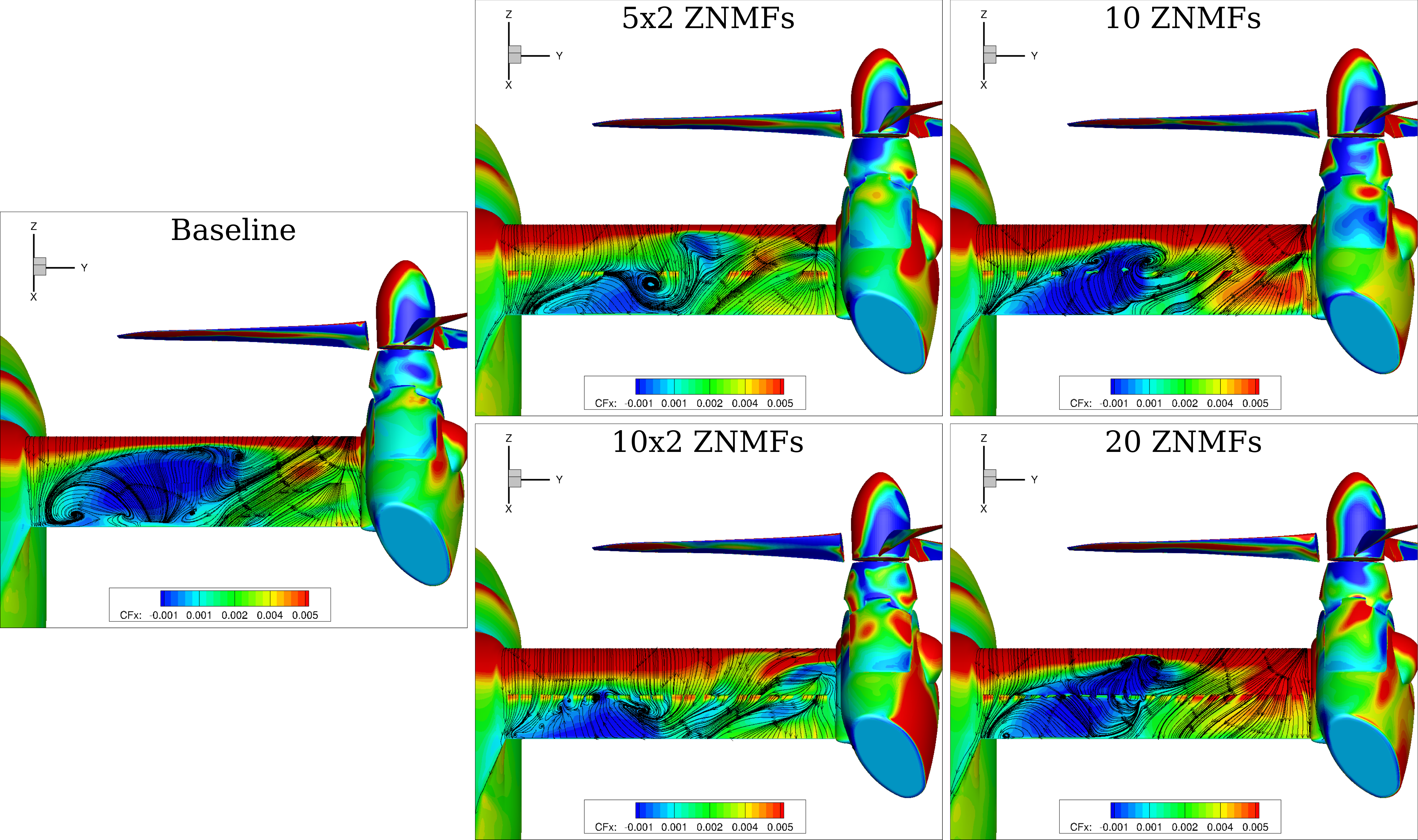}
\caption{Instantaneous skin friction $CF_x$ component contours with streamlines of CF ($CF_x, CF_y, CF_z$) on the aircraft surface for baseline (left) and the controlled cases with 5 pairs (top middle), 10 pairs (bottom middle), 10 (top right) and 20 (bottom right) ZNMF jets 
integrated on the aircraft with a tilted nacelle and rotating blades.}
\label{fig:case2_CFx_t0428_number}
\end{figure}

In previous section, the number of ZNMF devices necessary for active flow control was determined based on the results obtained for the baseline flow. It was found that by placing the ZNMF devices in position 1 an increase in aerodynamic efficiently was obtained, but placing the ZNMF devices in the other positions decreased the aerodynamic efficiency. 
One of the reasons could be that the number of ZNMF devices used was not sufficient to enhance the momentum of the boundary layer hence the use of active flow control was less efficient. In this section the influence of the number of ZNMF devices will be investigated. A straightforward approach is proposed by simply putting the ZNMF jets along the wing span from the fuselage to the nacelle. In total, four configurations were tested, i.e. 10 ZNMF jets, 20 ZNMF jets, 5 pairs of ZNMFs and 10 pairs of ZNMF, as can be seen in Fig.~\ref{fig:case2_CFx_t0428_number}. The instantaneous skin friction $CF_x$ component contours on the whole aircraft shows the flow separation zone in the baseline case occurring in the middle of the wing and it is transported toward the fuselage and the trailing edge due to the strong impact of the propeller. When active flow control using ZNMF jets is employed, the separation zone does not seem to be significantly reduced. This can be explained by the fact that all the ZNMF devices are distributed evenly on the wing and not concentrated in the middle of the wing where the vortices starting to form. Another cause is that these ZNMF devices should be shifted upstream closer to the leading edge of the wing.

\begin{figure}[h!]
\centering
\includegraphics[width=1.\textwidth]{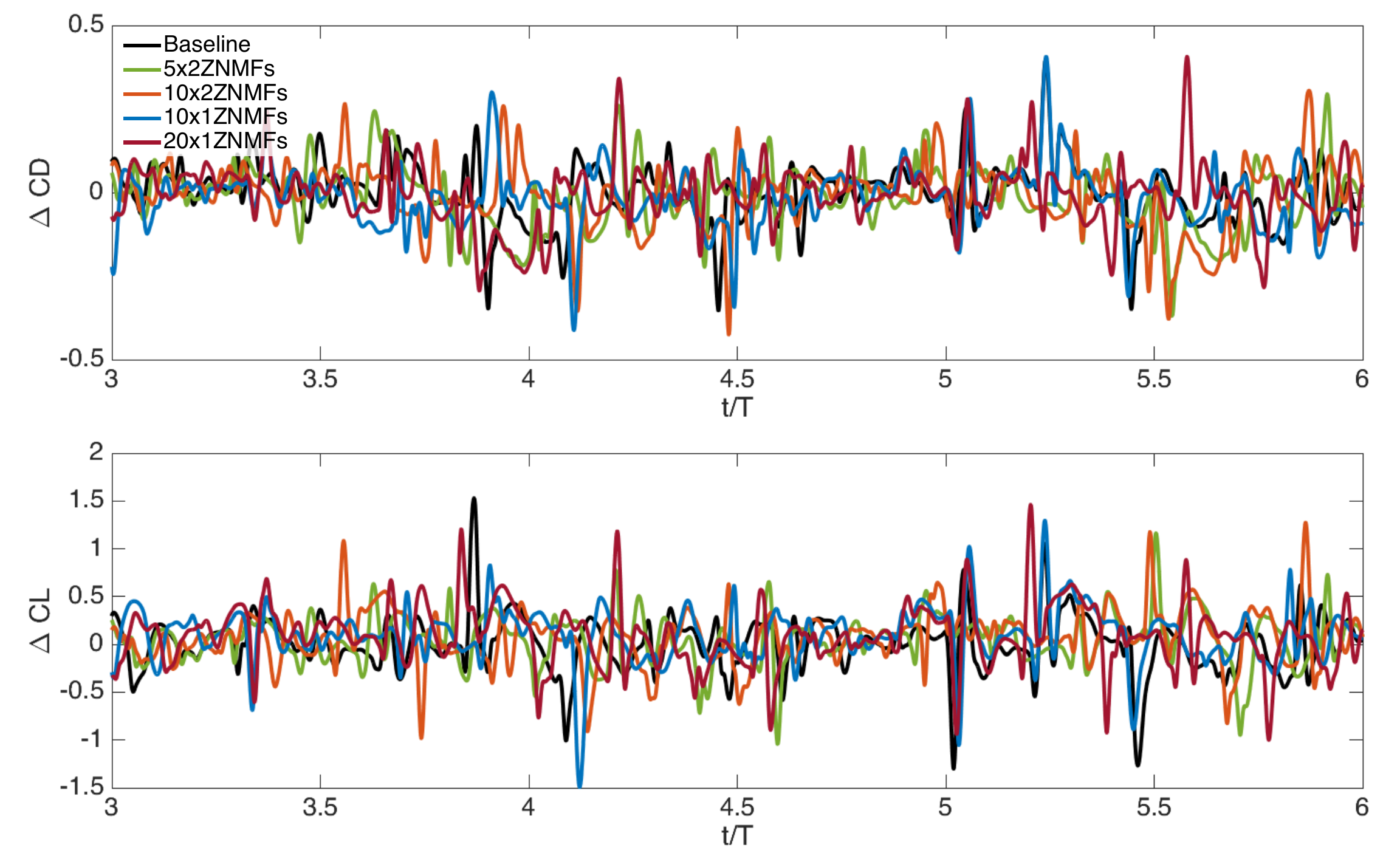}
\caption{Time history of the variation of the drag coefficient $\Delta CD$ (top) and the lift coefficient $\Delta CL$ (bottom) acting on the wing for different numbers of ZNMF with respect to the time-average forces of the baseline flow.}
\label{fig:case2_forces_number}
\end{figure}

\begin{figure}[h!]
\centering
\includegraphics[width=1.\textwidth]{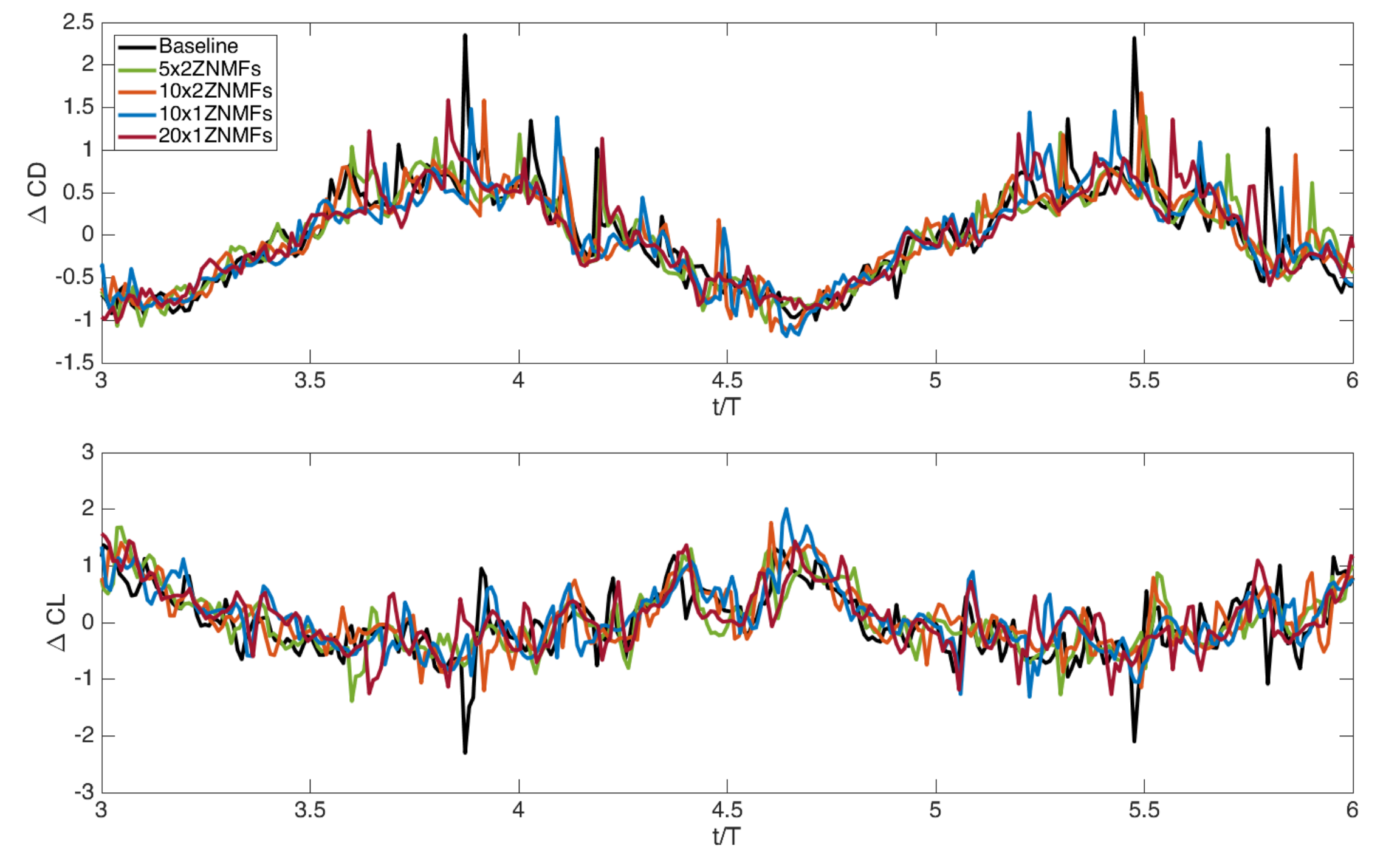}
\caption{Time history of the variation of the drag coefficient $\Delta CD$ (top) and the lift coefficient $\Delta CL$ (bottom) acting on the wing in the configuration 2 for different positions of the ZNMF with respect to the time-average forces of the baseline flow.}
\label{fig:case2_forces_number_aircraft}
\end{figure}

The variation of the time-average aerodynamic forces are calculated for the wing as well as on the whole aircraft, and are shown in Fig.~\ref{fig:case2_forces_number} and Fig.~\ref{fig:case2_forces_number_aircraft}, respectively. In general, the forces in the controlled case resemble the one in the baseline case, except for the reduction in some peaks of the baseline case. More particularly, the differences at around $3.9 t/T$ and $5.5 t/T$ are the most noticeable. Nevertheless, based on the time-average values summarized in Table~\ref{tab:forces_number}, the employment of ZNMF helps to improve significantly the aerodynamic performance through a reduction in drag and an increase in lift. The case with 10 ZNMFs distributed evenly on the wing appears to be the best configuration with a dramatic increase of lift-to-drag ratio up to $33.9\%$ for the wing and $8.82\%$ for the aircraft. Obviously, the lift and drag variation are less important when it is calculated for the whole aircraft. In addition, the drag reduction obtained on the wing is much larger comparing to the one on the aircraft since its form drag is considerably smaller than other components.

\begin{table}[h]
\begin{center}
\begin{minipage}{\textwidth}
\caption{The variation of the time-averaged force coefficients acting on the wing and the aircraft with respect to the baseline case for different numbers of ZNMF. The values are integrated for the 4th, 5th and 6th rotations of the propellers.}\label{tab:forces_number}
\begin{tabular*}{\textwidth}{@{\extracolsep{\fill}}lcccccc@{\extracolsep{\fill}}}
\toprule%
& \multicolumn{3}{@{}c@{}}{Wing} & \multicolumn{3}{@{}c@{}}{Aircraft} \\\cmidrule{2-4}\cmidrule{5-7}%
           & $\Delta C_D$ & $\Delta C_L$ & $\Delta (C_L/C_D)$ & $\Delta C_D$ & $\Delta C_L$ & $\Delta (C_L/C_D)$ \\
\midrule
5x2 ZNMF  & -15.3\% & 3.75\% & 22.49\% & -3.54\% & 1.92\% & 5.66\%\\
10x2 ZNMF  & -15.65\% & 6.33\%  & 26.06\%  & -3.5\% & 4.61\% & 8.41\%\\
10 ZNMF  & -19.72\% & 7.5\%  & 33.9\%  & -2.89\% & 5.67\% & 8.82\%\\
20 ZNMF  & -7.33\% & 6.96\%  & 15.42\%  & -1.98\% & 4.14\% & 6.24\%\\
\toprule
\end{tabular*}
\end{minipage}
\end{center}
\end{table}

\begin{figure}[hbt!]
\centering
\includegraphics[width=1.\textwidth]{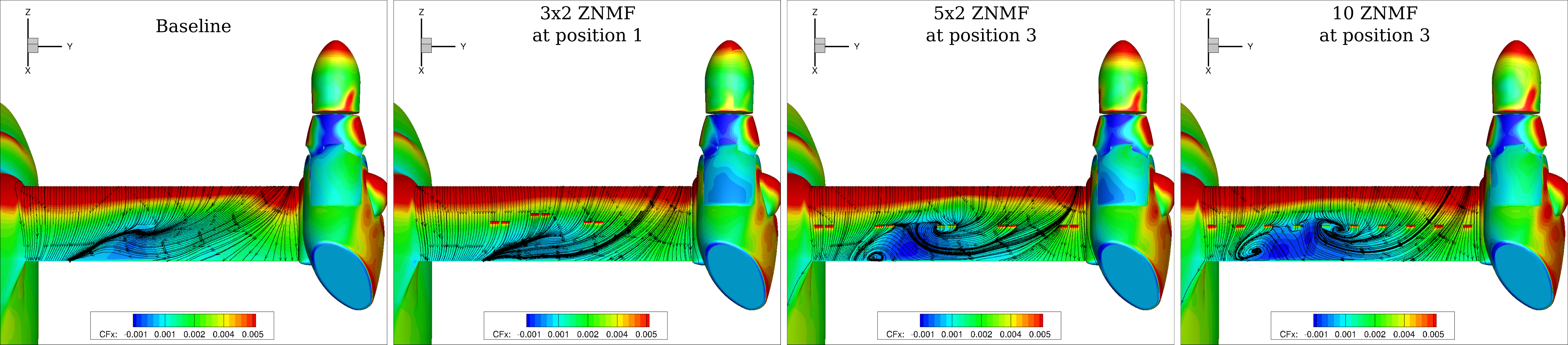}
\caption{Time-averaged skin friction $CF_x$ component contours with streamlines of $CF$ ($CF_x, CF_y, CF_z$) for the baseline and the controlled case with 3 pairs of ZNMFs at position 1, 5 pairs of ZNMF and 10 ZNMF. The blades of the propeller moving in time are removed for the calculation.}
\label{fig:case2_CF_baseline_vs_10ZNMFs}
\end{figure}

In order to understand the physics behind the benefit of using ZNMF devices for flow control, the time-averaged skin friction $CF_x$ is integrated and shown in Fig.~\ref{fig:case2_CF_baseline_vs_10ZNMFs}. Due to computational cost, only the baseline and the controlled case with 3 pairs of ZNMF at position 1, 5 pairs of ZNMF and 10 ZNMF jets have been calculated. Comparing to the instantaneous flow field presented earlier, the time-average skin friction does not help us to better understand the observed flow behavior.

\begin{figure}[hbt!]
\centering
\includegraphics[width=1.0\textwidth]{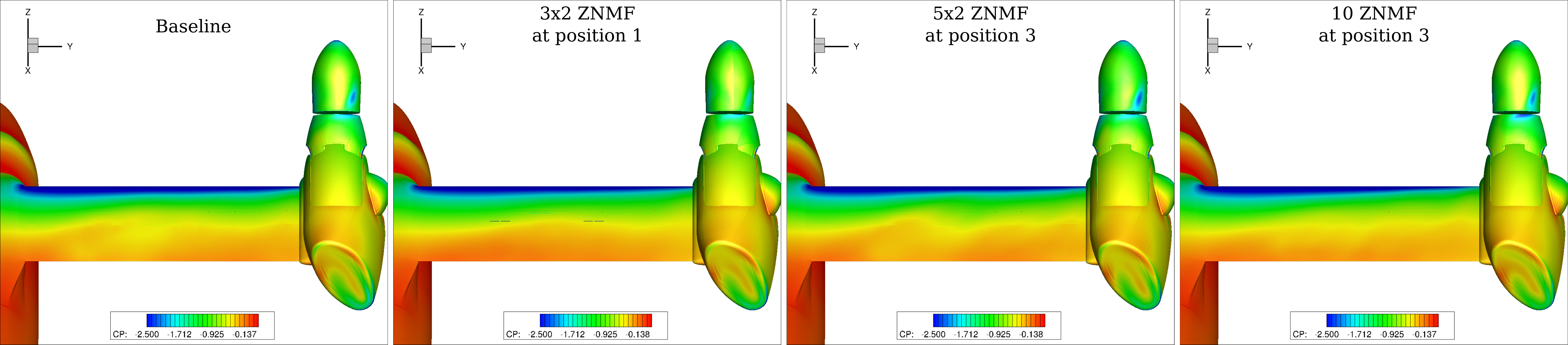}
\caption{Time-averaged pressure coefficient on the aircraft surface for the baseline and the controlled case with 3 pairs of ZNMFs at position 1, 5 pairs of ZNMF and 10 ZNMF. The blades of the propeller moving in time are removed for the calculation.}
\label{fig:case2_CP_baseline_vs_10ZNMFs}
\end{figure}

\begin{figure}[hbt!]
\centering
\includegraphics[width=1.0\textwidth]{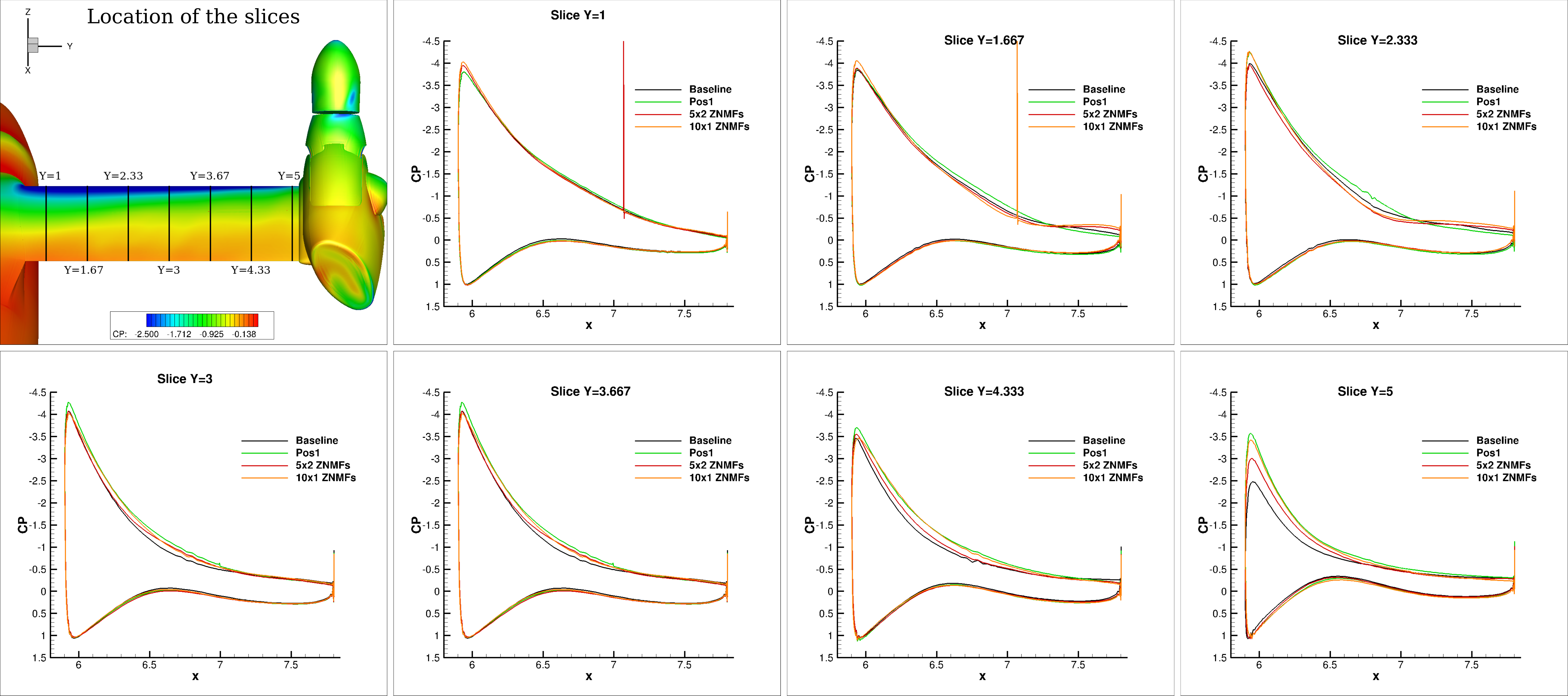}
\caption{Time-averaged pressure coefficient $CP$ distribution along the chordwise direction at different locations for the baseline and the controlled case with 3 pairs of ZNMFs at position 1 (green), 5 pairs of ZNMFs (red) and 10 ZNMFs (orange) at position 3.}
\label{fig:case2_CP_slices_baseline_vs_10ZNMFs}
\end{figure}

To better understand the increase in lift, the time averaged $CP$ distribution on the wing was also computed for these four cases (cf.~Fig.~\ref{fig:case2_CP_baseline_vs_10ZNMFs}). The results are quite similar at the wing leading edge, and in the wing-fuselage and wing-nacelle junction. Differences can be observed in the middle of the wing, where the $CP$ computed for the case of 3x2 ZNMFs in position 1 has a higher and more continuous behaviour. This was further analyzed by extraction of the averaged $CP$ distribution at several cross sections along the wing span (cf.~Fig.~\ref{fig:case2_CP_slices_baseline_vs_10ZNMFs}). When the ZNMF devices are used, the $CP$ on the extrados is smaller than the one in the baseline case. This can be the reason for the increase in lift. In addition, the ZNMF devices appear to effect also the flow upstream with a small low-pressure zone at the leading edge. Near the fuselage, the pressure at the leading edge in the case of 10x1 ZNMF jets is the smallest. Nevertheless, this is changed when going towards the nacelle. The $CP$ at the leading edge for the case of 3 pairs of ZNMF devices at position 1 is reducing much more than other cases near the nacelle. This can be responsible for the difference of pressure drag between the different configurations.

\section{Conclusions and Future Work}

 We presented here the study of active flow control using so called Zero-Net-Mass-Flux devices for the NGCTR aircraft, the Leonardo Next Generation of Civil Tilt Rotor. Numerical methods were employed using the NSMB CFD solver employing the chimera grid method to facilitate the modification of the nacelle tilting angle. Different numerical parameters were studied and in particular the effect of number of inner iterations of the dual time stepping method for unsteady simulations. Then the positions and numbers of the ZNMF devices were investigated for a critical configuration with a tilted nacelle of the full NGCTR aircraft. We focused firstly on the optimization of the position of 3 pairs of ZNMF devices integrated on the wing where the installation of 6 ZNMF devices in the middle of the wing resulted in a significant reduction in flow separation. Amongst the 5 ZNMF positions studied, placing these devices in the most upstream yielded the best performance with an increase in lift of the wing of 13.51\%. The second strategy concentrates on investigating the correlation between the number of employed ZNMF devices and their effectiveness. The study showed that using more ZNMF devices is not necessarily more efficient. More specifically, the flow control with 10 jets allows a slightly better aerodynamic performance comparing to the utilization of 20 ZNMF devices.

The results of these simulations demonstrated the benefits of ZNMF devices for active flow control for the NGCTR aircraft. However, the effectiveness of these devices is highly dependent on the location and the number of jets used for flow control. Further flow analysis and optimizations will be needed to identify the optimal parameters and locations of the ZNMF devices.

\section*{Acknowledgments}

The results presented in this paper are carried in the CleanSky2 project AFC4TR (funded by the European Union H2020 program under Grant Agreement 886718) and the H2020 project SMS (under Grant Agreement 723402). This work was granted access to the HPC resources of [CINES/TGCC] under the allocation 2020-2021-2022-[A0102A11355] made by GENCI. The authors would like to acknowledge the High Performance Computing Center of the University of Strasbourg for supporting this work by providing scientific support and access to computing resources. Part of the computing resources were funded by the Equipex Equip@Meso project (Programme Investissements d'Avenir) and the CPER Alsacalcul/Big Data. Leonardo Helicopters is acknowledged for providing the geometry, calculation conditions and for their support in performing the simulations.

\bibliography{sample}

\end{document}